\documentclass[twocolumn]{jpsj3}

\usepackage{graphicx,color}
\usepackage{amsmath}
\usepackage{amssymb}
\usepackage{bm}
\usepackage{txfonts}

\begin{document} 
\title{Field-angle-resolved Landscape of Non-Fermi-liquid Behavior\\ in the Quasi-kagome Kondo Lattice CeRhSn}

\author{
	Shunichiro \textsc{Kittaka},$^{1,2}$\thanks{kittaka@phys.chuo-u.ac.jp}
	Yohei \textsc{Kono},$^{1,2}$
	Suguru \textsc{Tsuda},$^3$\\
	Toshiro \textsc{Takabatake},$^3$ and
	Toshiro \textsc{Sakakibara}$^{2}$ 
}

\inst{$^{1}$Department of Physics, Faculty of Science and Engineering, Chuo University, Kasuga, Bunkyo-ku, Tokyo 112-8551, Japan\\
      $^{2}$Institute for Solid State Physics, University of Tokyo, Kashiwa, Chiba 277-8581, Japan\\
      $^{3}$Quantum Matter Program, Graduate School of Advanced Science and Engineering, Hiroshima University, Higashi-Hiroshima 739-8530, Japan
}

\date{\today}

\abst{
We have employed a magnetic field angle as a tuning parameter 
in a comprehensive measurement of the specific heat, magnetocaloric effect, and magnetization for the quasi-kagome Kondo lattice CeRhSn, 
which is considered to exhibit zero-field quantum criticality driven by geometrical frustration.
By constructing the field-angle-resolved landscape of the entropy, 
we unexpectedly revealed that the non-Fermi-liquid nature survives up to a metamagnetic crossover field of roughly 3~T 
in the very narrow field-orientation range, close to the direction parallel to the quasi-kagome plane.
We propose that spin fluctuations along the hexagonal $c$ axis are the dominant driving force for the non-Fermi-liquid behavior
because it is strongly suppressed by a magnetic-field component along the $c$ axis.
The multidimensional entropy landscape, which directly reflects the degeneracy of ground states, opens a new route for uncovering the nature of exotic phases in anisotropic systems.
}

\maketitle

\section{Introduction}

CeRhSn has a hexagonal ZrNiAl-type structure with space group P$\bar{6}$2m,
in which basal layers are stacked alternately along the $c$ axis and 
the Ce sublattice can be viewed as a quasi-kagome lattice [inset of Fig. \ref{angle}(a)].
Such an arrangement allows geometrical frustration if the nearest-neighbor Ising spins are coupled antiferromagnetically.
Isostructural compounds CePdAl, YbAgGe, and UCoAl show a variety of metamagnetic transitions induced by geometrical frustration \cite{Mochidzuki2017JPSJ,Tokiwa2013PRL} or a ferromagnetic quantum critical point (QCP).\cite{Aoki2011JPSJ}
For CeRhSn, the magnetic susceptibility increases drastically upon cooling under a magnetic field $H$ along the $c$ axis,
although local $4f$ electrons of Ce ions are coupled to conduction electrons below the Kondo temperature ($\sim 200$~K).\cite{Kim2003PRB}
This indicates that local moments are not fully screened even at low temperatures.
Thereby, the magnetic susceptibility in $H \parallel c$ is roughly 30 times larger than that in $H \parallel a$.
The reduced moments were estimated to be 0.05 $\mu_{\rm B}$/Ce from the magnetization curve in $H \parallel c$ ($\mu_{\rm B}$ is the Bohr magneton).\cite{Kim2003PRB}
The presence of antiferromagnetic interactions was suggested 
by the negative values of the Curie-Weiss temperature of $-390$ K for $H \parallel  a$ ($-57$ K for $H \parallel c$)\cite{Kim2003PRB} and
by the results of $^{119}$Sn-NMR experiments on powder samples.\cite{Tou2004PRB}
In zero field, however, no long-range magnetic order has been found at least above 50 mK.\cite{Schenck2004JPSJ}
Instead, non-Fermi-liquid (NFL) behaviors were observed in the magnetic susceptibility, the electrical resistivity, and the specific heat at low temperatures.\cite{Kim2003PRB}
The Gr\"{u}neisen ratio $\Gamma(T)$, determined from the volume thermal expansion and specific heat, shows 
the power-law temperature dependence as $\Gamma(T)=aT^{-1.6} + b$
which is a general hallmark of quantum criticality.

Recently, it was reported that the NFL nature of CeRhSn was fragile under a finite magnetic field;\cite{Tokiwa2015SA}
the NFL behaviors were rapidly suppressed with increasing field in any direction.
This fact was attributed to the presence of a QCP near $H=0$. 
The specific heat coefficient and the magnetic Gr\"{u}neisen ratio $\Gamma_H=1/T(\partial T/\partial H)|_S$, 
both of which diverge on cooling in zero field, are strongly suppressed by increasing $H$.
Furthermore, the thermal expansion $\alpha$ exhibits the anisotropic temperature dependence and shows a divergent behavior in $\alpha/T$ only along the $a$ axis. 
This fact suggests that the NFL nature is coupled to in-plane uniaxial pressure, which breaks the symmetry of the quasi-kagome lattice.
By applying uniaxial pressure along the $a$ axis, the NFL behavior was indeed destroyed and a complicated magnetic phase diagram was induced.\cite{Kuchler2017PRB}
On the basis of these facts, CeRhSn has been suggested to be a promising frustrated Kondo-lattice metal 
with a quantum spin-liquid ground state.\cite{Tokiwa2015SA,Vojta2018RPP,Balents2010Nature}

Another fascinating character of CeRhSn is the occurrence of a metamagnetic crossover under a magnetic field parallel to the $a$ axis.\cite{Tokiwa2015SA,Yang2017PRB}
Recently, a similar metamagnetic feature was reported for the isostructural CeIrSn ($T_{\rm K}=480$~K) in $H \parallel a$ as well.\cite{Tsuda2018PRB}
In the previous report,\cite{Tokiwa2015SA} the origin of this metamagnetic crossover was discussed in the analogy of the isostructural antiferromagnet YbAgGe,
which exhibits spin-flop bicriticality under a magnetic field along the easy-magnetization plane ($\parallel ab$).\cite{Tokiwa2013PRL}
It has been speculated that, due to the suppression of the magnetic-ordering temperature, 
the spin-flop metamagnetic transition becomes a crossover in CeRhSn. 

As introduced above, the quasi-kagome Kondo-lattice CeRhSn exhibits highly anisotropic properties with unusual quantum critical phenomena. 
In such highly-anisotropic materials, the field-orientation effect needs to be carefully investigated 
because a tiny component of an easy-axis magnetic field may cause dominant effect on its physical properties.
In this study, to uncover which kind of spin fluctuations plays key roles in arising the strange metallic behavior in CeRhSn, 
field-angle-resolved measurements of its specific heat $C$, magnetocaloric effect, and magnetization $M$ have been performed at low temperatures.

\section{Methods}
Single crystals of CeRhSn were grown by the Czochralski method using an rf induction furnace.\cite{Kim2003PRB}
A single piece with a weight of 44.0 mg mass was used in this study.
It was cut into a thin slab, whose dimensions were roughly 3 mm $\times$ 3 mm $\times$ 0.2 mm 
along the $a$, $c$, and $a^\ast$ axes, respectively.
The residual resistivity ratio of CeRhSn is known to be at most 4.~\cite{Kim2003PRB}

The dc magnetization was measured using a home-built capacitively detected Faraday magnetometer \cite{Sakakibara1994JJAP} combined with a two-axis alignment device,\cite{Nakamura2018JPSJ} 
which was installed in a dilution refrigerator (Kelvinox MX100, Oxford). 
The magnetic field was generated along the vertical $z$ direction using a solenoid magnet equipped with a gradient coil producing the field gradient in the $z$ direction ($dH_z/dz$). 
The field gradient of 1 (5) T/m was used for the measurement at 0.15 (4.2) K.
The sample was mounted on the capacitor transducer so that the $a$ ($c$) axis was oriented close to the $z$ ($x$) direction. 
The angle between the field and the $ab$ plane $\phi_H$ was tuned by rotating the transducer around the $y$ axis using a home-made tilting stage.
A magnetic force $F_z$ proportional to the sample magnetization $M_z$, $F_z=M_zdH_z/dz$, in addition to the magnetic-torque background, 
was detected as a capacitance change of the capacitor transducer.
The magnetic-torque component was measured separately under a zero field gradient $dH_z/dz=0$, and its contribution was subtracted from the data.
In the magnetization measurements, $\phi_H=0^\circ$ was defined as the angle at which the magnetic-torque component is almost eliminated.\cite{SM}
For comparison, the magnetization was also measured using a commercial SQUID magnetometer (MPMS, Quantum Design). 

The specific heat and the field-rotational magnetocaloric effect were measured by the quasi-adiabatic technique using a home-built calorimeter.\cite{Kittaka2018JPSJ}
The field-rotational magnetocaloric effect is the observed temperature change in response to the adiabatic rotation of the external magnetic field.
In the present study, this effect was investigated by fitting the initial slope of the temperature change, $dT/d\phi_H$, during each field rotation by $1^\circ$ or $0.5^\circ$.
Using the thermodynamic relation, $(\partial S/\partial \phi_H)_{T,H}=-C/T(\partial T/\partial \phi_H)_{S,H}$,
the relative entropy change, $\Delta S_\phi(H,\phi_H)=S(H,\phi_H)-S(H,90^\circ)$, can be evaluated as $\Delta S_\phi=-\int^{\phi_H}_{90^\circ} C/T(dT/d\phi_H)d\phi_H$.
Likewise, the field variance of the entropy $\Delta S_H(H,\phi_H)=S(H,\phi_H)-S(H=0)$ was evaluated as $\Delta S_H=-\int^{H}_{0} C/T(dT/dH)dH$ 
using the relation $(\partial S/\partial H)_{T,\phi_H}=-C/T(\partial T/\partial H)_{S,\phi_H}$, 
i.e., the conventional field-sweep magnetocaloric effect.
In these thermal measurements, the sample was mounted on addenda of the calorimeter so that the $a^\ast$ axis was oriented along the $z$ direction.
It was cooled in a dilution refrigerator (Kelvinox AST Minisorb, Oxford) whose base temperature was well below 60~mK.
The magnetic field was generated in the horizontal $x$ direction up to 5~T using a split-pair magnet.
By rotating the refrigerator around the $z$ axis ($\parallel a^\ast$) using a stepper motor,
the magnetic field was rotated on the $ac$ plane.

The $\Delta S_\phi$ data at different $H$ were determined by adding an offset of $\Delta S_H(H,90^\circ)$ for each curve.
In this way, we obtained the entropy change relative to $S(H=0)$, i.e., $\Delta S_\phi(H,\phi_H)+\Delta S_H(H,90^\circ)=S(H,\phi_H)-S(H=0)$, 
which is referred to here as $\Delta S(H,\phi_H)$.
Using $S=\gamma T$  at 5 T for $H \parallel c$ and $\Delta S_H(5\ \rm{T}, 90^\circ)=-21.5$~mJ/(mol K) at 0.1 K, 
both of which are obtained in the present study [Fig. 3(b)],
the absolute value of $S$ was calibrated in the overall $T$--$H_{\parallel a}$--$H_{\parallel c}$ space.

\section{Results}
\subsection{Magnetization measurements}

\begin{figure}[t]
\begin{center}
\includegraphics[width=3.1in]{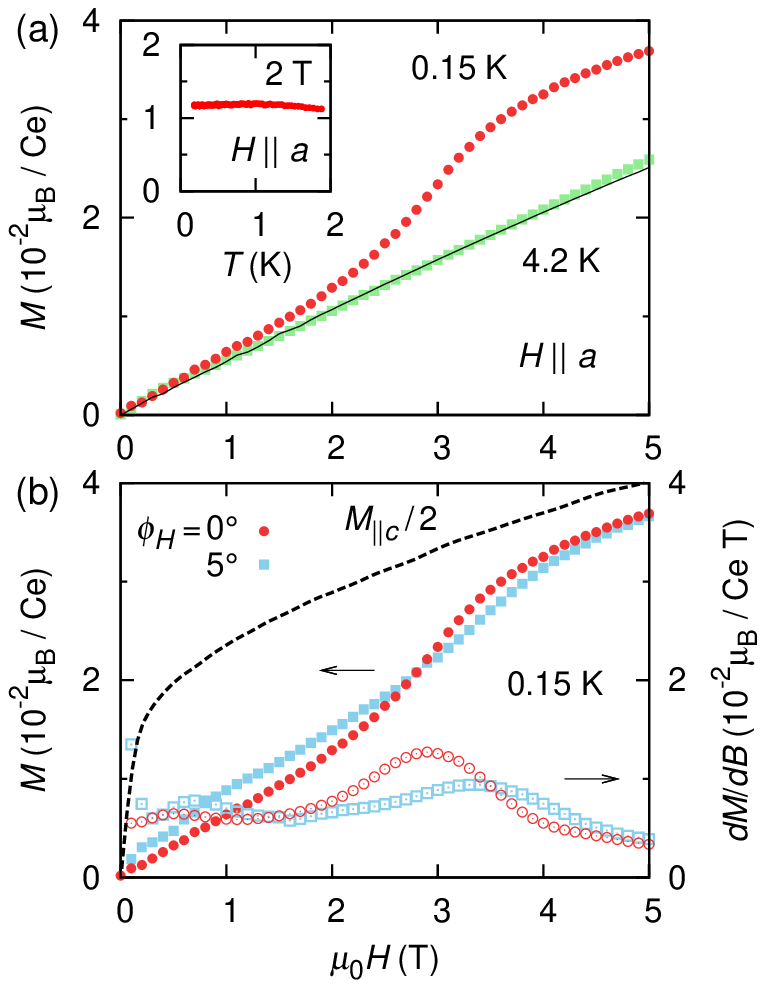} 
\end{center}
\caption{
(Color online) 
(a) Field dependence of the magnetization $M(H)$ at 0.15 (circles) and 4.2~K (squares) in the exact in-plane direction.
The solid line represents $M(H)$ at 4.2~K measured by MPMS in the approximate in-plane direction. 
Inset: $M(T)$ at 2~T in $H \parallel a$.
(b) $M(H)$ at 0.15~K for $\phi_H = 0^\circ$ (closed circles) and $5^\circ$ (closed squares).
Open symbols represent the field derivative of the magnetization.
The dashed line shows the previously reported $M(H)$ curve at 0.5~K in $H \parallel c$,\cite{Kim2003PRB} which is divided in half.
}
\label{MH}
\end{figure}

To obtain direct information on the magnetic property of CeRhSn under an exact in-plane magnetic field,
dc magnetization was measured. 
The angle between the magnetic field and the $ab$ plane, $\phi_H$, was carefully determined within an error of better than $1^\circ$ from magnetic-torque measurements by using the same setup.\cite{SM}
Symbols in Fig.~\ref{MH}(a) show the magnetization curve, $M(H)$, measured at 0.15 and 4.2 K for $\phi_H=0^\circ$.
At 0.15 K, $M(H)$ shows an abrupt increase around $\mu_0H_{\rm m} \sim 3$~T, 
providing direct evidence for the occurrence of a metamagnetic crossover.
By increasing temperature up to 4.2~K, $M(H)$ shows only a Pauli paramagnetic contribution; the metamagnetic crossover disappears in this field range. 
At 4.2 K, the $M(H)$ curve was also measured by using MPMS and is represented by a solid line in Fig.~\ref{MH}(a).
The excellent agreement between the $M(H)$ curves taken with the two instruments ensures high reliability of our magnetization data.

It is noted that the $M(H)$ curves at 0.15 and 4.2~K are roughly the same up to 2~T, reminiscent of the FL nature. 
In fact, $M(T)$ at 2~T in $H \parallel a$ very weakly depends on temperature below 2~K [inset of Fig.~\ref{MH}(a)].
As discussed later, in the case of CeRhSn, the NFL nature is not detected from $M(T)$ due to the thermodynamic Maxwell relation.
When the field orientation is tilted away from the $ab$ plane by $5^\circ$, $H_{\rm m}$ increases slightly, and the low-field $M(H)$ is enhanced.
The former can be more clearly seen in the field derivative of the magnetization, i.e., $dM/dB$ [Fig.~\ref{MH}(b)].
The latter can be attributed to the $c$-axis component of magnetization.

\subsection{Calorimetric measurements}
\begin{figure}[t]
\begin{center}
\includegraphics[width=3.1in]{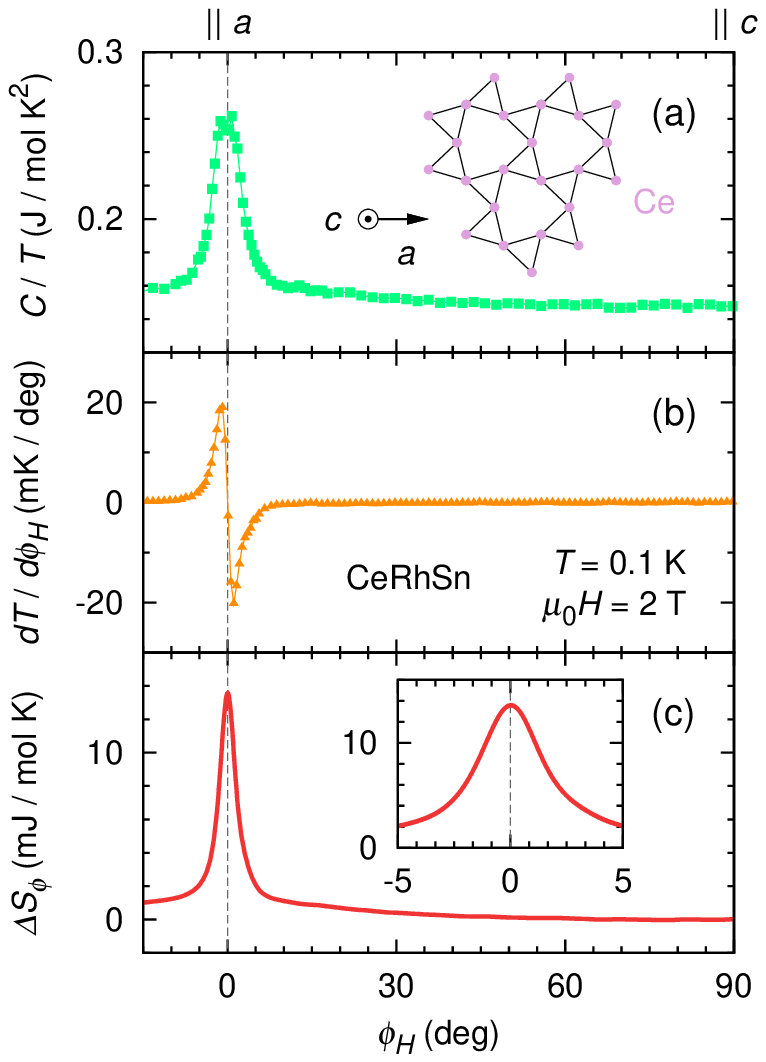}
\end{center}
\caption{
(Color online) 
(a) The specific heat divided by temperature $C/T$, (b) field-rotational magnetocaloric effect $dT/d\phi_H$, and (c) relative entropy change $\Delta S_\phi$ as a function of $\phi_H$ at 0.1~K.
Here, the applied magnetic field of 2~T is rotated within the $ac$ plane.
Inset in (a): top view along the $c$ axis of the quasi-kagome lattice formed by Ce atoms in CeRhSn.
Inset in (c): enlarged plot of $\Delta S_\phi$ near $\phi_H=0^\circ$.
}
\label{angle}
\end{figure}

To apply the magnetic field exactly parallel to the $ab$ plane,
the field-angle $\phi_H$ dependence of the specific heat $C$ was investigated at 0.1~K by rotating a magnetic field of 2~T within the $ac$ plane.
As shown in Fig.~\ref{angle}(a), a sharp peak was observed in the $C(\phi_H)/T$ data centered at $\phi_H=0^\circ$.
Nearly the same $\phi_H$ dependence of the specific heat was obtained from a measurement in a rotating field within the $a^\ast c$ plane, perpendicular to the $a$ axis. 
These results indicate that the enhancement of the specific heat under an in-plane magnetic field can be easily destroyed by field misalignment of only a few degrees.

To provide further evidence for the anomalous state under a precisely-aligned in-plane magnetic field, 
a relative change in the entropy $S$ with $\phi_H$ was investigated via the field-rotational technique.\cite{Kittaka2018JPSJ}
Figure \ref{angle}(b) shows the field-rotational magnetocaloric effect at 0.1~K in 2~T within the $ac$ plane.
By using the thermodynamic relation, $(\partial S/\partial \phi_H)_{T,H}=-C/T(\partial T/\partial \phi_H)_{S,H}$,
the relative change in $S$ with $\phi_H$, $\Delta S_\phi=S(\phi_H)-S(90^\circ)$, was calculated and plotted in Fig.~\ref{angle}(c).
$\Delta S_\phi$ shows a very sharp peak at $\phi_H=0^\circ$ with a full width at half maximum of at most $4^\circ$ [inset of Fig.~\ref{angle}(c)]. 
This is evidence that the NFL behavior is robust against $H$ up to 2~T only when $H$ is exactly parallel to the quasi-kagome plane.

\begin{figure}[t]
\begin{center}
\includegraphics[width=3.1in]{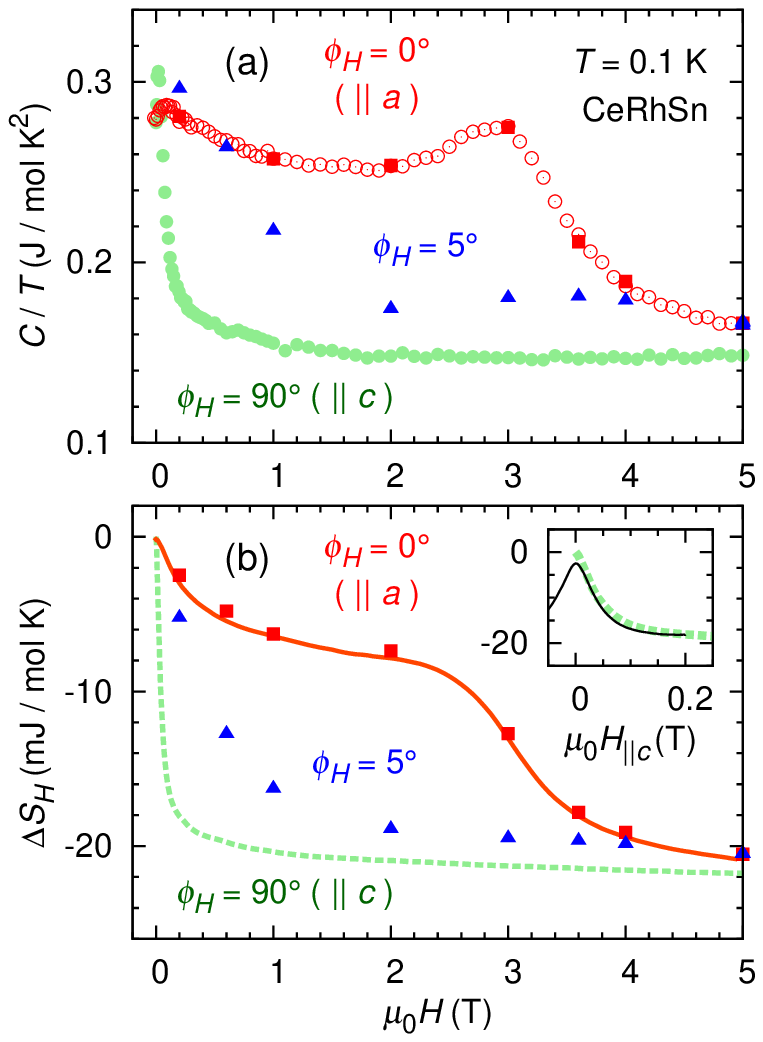} 
\end{center}
\caption{
(Color online) 
Field dependence of (a) $C(H)/T$ and (b) $\Delta S_H$ at 0.1~K for $\phi_H=0^\circ$ and $90^\circ$.
The solid squares and triangles in (a) [(b)] are the field-rotational $C(\phi_H)/T$ [$\Delta S(H,\phi_H)$] data at $\phi_H=0^\circ$ and $5^\circ$, respectively.
Inset in (b): $\Delta S_H(H,90^\circ)$ (dashed line) and $\Delta S(0.2\ {\rm T},\phi_H)$ (solid line) at 0.1~K as a function of $H_{\parallel c}$. 
}
\label{Field}
\end{figure}

However, when the in-plane field exceeds approximately 3~T, the high-entropy state is destroyed abruptly.
Figures~\ref{Field}(a) and \ref{Field}(b) show the field variations of $C/T$ and the entropy change $\Delta S_H=S(H)-S(0)$ at 0.1 K, respectively. 
The latter is evaluated by using the relation $(\partial S/\partial H)_{T,\phi_H}=-C/T(\partial T/\partial H)_{S,\phi_H}$, i.e., the conventional field-sweep magnetocaloric effect.
In $H \parallel a$, the entropy is gradually released at low fields below roughly 1~T 
and decreases remarkably at $\mu_0H_{\rm m}\sim 3$~T [Fig.~\ref{Field}(b)].
Around $H_{\rm m}$, $C(H)/T$ shows an anomalous peak [Fig.~\ref{Field}(a)], similar to the one observed in the previous report (Fig. 3B in Ref. \ref{Tokiwa2015SA}).
An entropy plateau arises under an exact in-plane magnetic field ($\phi_H=0^\circ$) in the range $1\ {\rm T} \lesssim \mu_0H \lesssim 2$~T.

\begin{figure}[t]
\begin{center}
\includegraphics[width=3.1in]{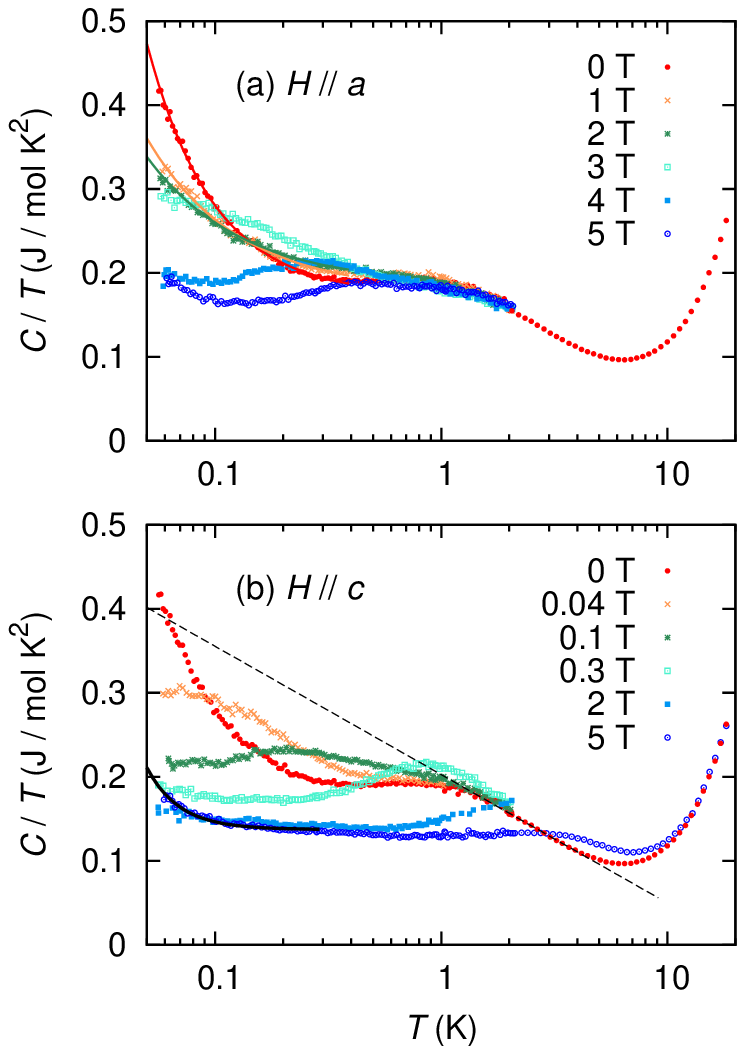} 
\end{center}
\caption{
(Color online) 
Temperature dependence of $C/T$ at (a) $\phi_H=0^\circ$ ($H \parallel a$) and (b) $\phi_H=90^\circ$ ($H \parallel c$).
The high-temperature data for $T>2$~K are taken from Ref. \ref{Higaki2006JPSJ}.
Solid lines in (a) represent the fits to the low-temperature data at 0, 1, and 2~T using a function $f(T)=\alpha T^{-n} + \gamma_0$.
The fitting parameters are shown in Table~\ref{Table}. 
The dashed line in (b) shows $-\ln T$ behavior suggested from the zero-field data.
The solid line in (b) indicates the nuclear specific-heat contribution.
}
\label{CT}
\end{figure}

Figures~\ref{CT}(a) and \ref{CT}(b) show the temperature dependence of $C/T$ at various fields for $\phi_H=0^\circ$ and $90^\circ$, respectively.
In zero field, $C/T$ increases in proportion to $-\ln T$ on cooling below 4~K,\cite{Higaki2006JPSJ,Kim2003PRB} 
although $T_{\rm K}$ ($\sim 200$~K) is much higher than the present temperature range.
This $-\ln T$ increase levels off below 2~K, apparently suggesting the formation of the Kondo-singlet state.
On further cooling, however, $C/T$ is again enhanced for $T<0.5$~K and exceeds the value of 0.4 J/(mol K$^2$) at $T<0.07$~K.
This zero-field behavior is in good agreement with the previous report.\cite{Tokiwa2015SA}
This fact suggests that the divergent behavior does not depend on the sample quality.
Under a magnetic field along the easy-magnetization $c$ axis [Fig.~\ref{CT}(b)], 
the low-temperature upturn in $C(T)/T$ shifts toward the higher temperature side and changes into a peak structure.
At 5~T in $H \parallel c$, both the peak structure and the enhancement of $C(T)/T$ around 1~K due to $-\ln T$ behavior are strongly suppressed, 
resulting in the FL behavior below 3~K.
Note that the weak upturn at low temperatures can be attributed to the nuclear contribution [$C \propto T^{-2}$; solid line in Fig. \ref{CT}(b)].

\begin{table}[t]
\begin{center}
\caption{
Parameters obtained by the fits to the $C(T)$ data for $H \parallel a$ below 0.4~K using a power-law function $f(T)=\alpha T^{-n} + \gamma_0$.
}
\vspace{0.07in}
\renewcommand{\arraystretch}{1.4}
\begin{tabular*}{8.8cm}{@{\extracolsep{\fill}}cccc}\hline\hline
$\mu_0H_{\parallel a}$ (T) & $n$ & $\alpha$ (mJ/mol K$^{2+n}$) & $\gamma_0$ (mJ/mol K$^2$) \\ \hline
0 & 1.48 & 3.56 & 173\\
1 & 1.08 & 7.21 & 177\\
2 & 1.04 & 6.75 & 186\\
\hline\hline
\label{Table}
\end{tabular*}
\end{center}
\end{table}

Under an exact in-plane magnetic field [Fig.~\ref{CT}(a)],
the low-temperature upturn in $C(T)/T$ is rather robust at least up to 2~T,
in sharp contrast to the previous observation.\cite{Tokiwa2015SA}
This upturn can be characterized by the power-law fit using the function, $f(T)=\alpha T^{-n}+\gamma_0$.
The parameters obtained by the fits are summarized in Table~\ref{Table}.
With increasing $H$ from 0 to 1~T, the exponent $n$ changes drastically from $n \approx 1.5$ to $n \approx 1.0$. 
However, it is nearly unchanged in the entropy-plateau regime ($n \approx 1.0$ for $1\ {\rm T} \lesssim \mu_0H \lesssim 2$~T). 
For $\mu_0H\gtrsim3$~T, the fit with $n \approx 1.0$ is not successful because of the appearance of the peak structure.
At 5~T, $C(T)/T$ in $H \parallel a$ exhibits the FL-like behavior below 0.2~K, 
although it remains to show a peak around 0.5~K and the enhancement of $C(T)/T$ around 1~K remains rigid.

One may suspect that the low-temperature upturn in $C(T)/T$ at 2~T is incompatible with the FL-like behavior in $M(T)$ for $H \parallel a$.
However, the plateau in $C(H)/T$ around 2~T [Fig.~\ref{Field}(a)] is consistent with the temperature-independent magnetization [inset of Fig.~\ref{MH}(a)], 
satisfying the thermodynamic Maxwell relation $\partial (C/T) / \partial H=\partial^2 M/\partial T^2$.
These contrasting temperature responses of $M(T)$ and $C(T)/T$, similar to the recent observations in Ni-doped CeCoIn$_5$,\cite{Yokoyama2019PRB}
indicate that the NFL nature is coupled to spin fluctuations along the $c$ axis, which do not affect the in-plane magnetization directly.

\subsection{Field-angle-resolved mapping of thermodynamic quantities}
\begin{figure*}[t]
\begin{center}
\includegraphics[width=6.2in]{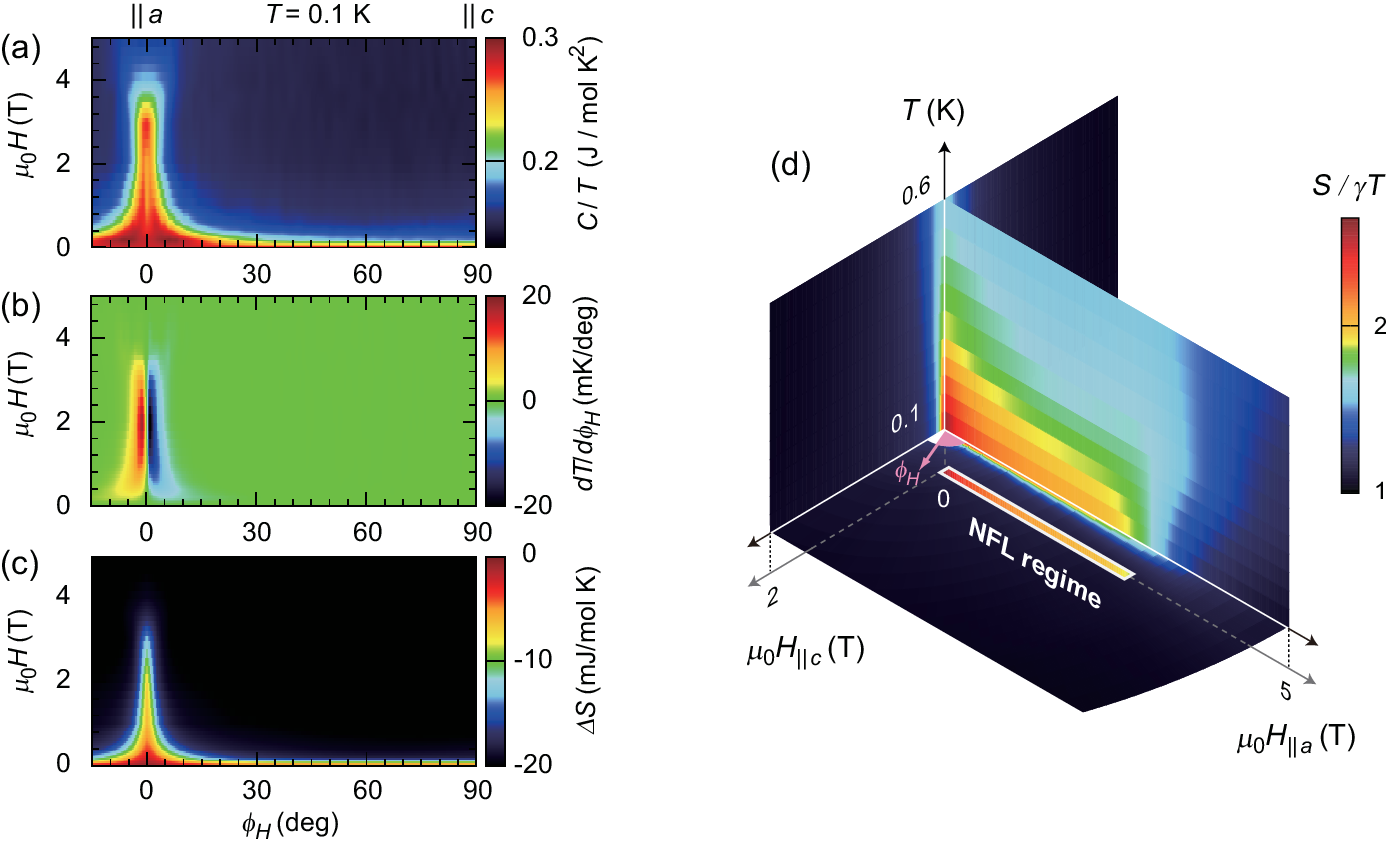} 
\end{center}
\caption{
(Color online) 
Contour plots of (a) $C/T$, (b) $dT/d\phi_H$, and (c) $\Delta S$ in the $H$--$\phi_H$ plane at 0.1~K.
(d) Entropy landscape in the $T$--$H_{\parallel a}$--$H_{\parallel c}$ space, where 
$S/\gamma T$ is depicted using color on three selected planes for $T\ge0.1$~K.
Here, the low-temperature value of $C/T$ at 5 T for $H \parallel c$ is identified as the Sommerfeld coefficient, i.e. $\gamma=132$~mJ/(mol K$^2$).
A NFL line at $H_{\parallel c}=0$ and $|\mu_0H_{\parallel a}| \lesssim 2.5$~T is represented by a thick line outlined in white.
See also Supplemental Material Figs. S4 and S5.
}
\label{phase}
\end{figure*}

Figures~\ref{phase}(a)--\ref{phase}(c) represent the contour maps of $C$, $dT/d\phi_H$, and $\Delta S$ [$=S(H,\phi_H)-S(H=0)$] in the $H$--$\phi_H$ plane at 0.1~K, 
which were constructed via field-rotational measurements.
The data points at $\phi=0^\circ$ extracted from Figs.~\ref{phase}(a) and \ref{phase}(c) are plotted in Figs.~\ref{Field}(a) and \ref{Field}(b) by solid squares 
for comparison, in good agreement with those obtained by the field-sweep measurements (open circles).  
Figure~\ref{phase}(d) summarizes the phase diagram in the $T$--$H_{\parallel a}$--$H_{\parallel c}$ space, where $H_{\parallel a}=H\cos\phi_H$ and $H_{\parallel c}=H\sin\phi_H$.
The high-entropy state arises in the narrow angle range for $|\phi_H|\lesssim 2^\circ$ 
until $H$ reaches $H_{\rm m}$ or $H_{\parallel c}$ exceeds a critical value.
It should be emphasized that, with increasing $H$, the low-temperature entropy is stepwisely (monotonically) released at $\phi_H=0^\circ$ ($\phi_H=90^\circ$) and becomes nearly isotropic above $H_{\rm m}$.
The extrapolation of the entropy landscape to the $H_{\parallel a}$--$H_{\parallel c}$ plane at 0~K demonstrates 
the existence of a "NFL" regime at $H_{\parallel c}=0$ and $|\mu_0H_{\parallel a}| \lesssim 2.5$~T.

At $\phi_H=5^\circ$, $S(H)$ decreases rapidly with increasing $H$ [triangles in Fig.~\ref{Field}(b)], well below $H_{\rm m}$ [see Fig.~\ref{MH}(b)].
This means that the NFL nature is more easily destroyed by $H_{\parallel c}$.
In fact, for $H\parallel c$, the NFL behavior is sensitively suppressed at very low fields below 0.1~T.
This monotonic and rapid decrease of the entropy by $H_{\parallel c}$ can also be confirmed from the $\Delta S_\phi$ data in low fields [solid line in the inset of Fig.~\ref{Field}(b)]. 
The good match between the solid and dashed lines in the inset of Fig.~\ref{Field}(b) demonstrates that 
$H_{\parallel c}$ is the field that is conjugate to the critical fluctuations.

\section{Discussion}
Let us discuss the origin of the NFL behavior in CeRhSn.
Key features found in this study are as follows.
First, the NFL behavior is highly anisotropic at low fields below 3~T. 
It is easily destroyed by the application of $H_{\parallel c}$, with a rapid suppression of the low-temperature entropy.
By contrast, no NFL behavior is observed in the in-plane magnetization in $H_{\parallel a}$.
Second, the NFL behavior of  $C/T\sim T^{-n}$, $n\approx 1$ is nearly invariant in the entropy-plateau regime $1\ {\rm T} \lesssim \mu_0H_{\parallel a} \lesssim 2$~T, 
followed by an increase of the exponent to $n\approx 1.5$ for $\mu_0H_{\parallel a}<1$~T.
Third, this strongly anisotropic NFL behavior is destroyed upon the metamagnetic crossover occurring at $\mu_0H_{\rm m}\sim 3$~T. 
Accordingly, the low-temperature entropy is stepwisely released in $H \parallel a$  and becomes nearly isotropic above $H_{\rm m}$.

We note that the NFL exponent of $n=1-1.5$ in $C/T$ is a transient value, and is not a critical exponent associated with a quantum criticality. 
This is because the integrated entropy diverges as $T\rightarrow 0$ when $n\geq 1$. 
Thus, at still lower temperatures, $C/T$ would either exhibit a true critical behavior with $n<1$ or level off to a FL state. 
To disclose the fate of the NFL behavior in CeRhSn, measurements below 50~mK will be needed.
It is also noteworthy that the $c$-axis susceptibility increases as $\chi_c (T) \propto T^{-1.1}$ below $\sim 2$~K down to 0.4~K.\cite{Kim2003PRB}
In this regard, we would like to refer to the previous $\mu$SR experiment,\cite{Schenck2004JPSJ}
which claims that CeRhSn shows a separation into Ce$^{3+}$ domains and domains in which Ce undergoes valence fluctuations, 
and the Ce$^{3+}$ domains occupy an increasing fraction of the sample volume below 1.3~K and finally involve the whole volume below $\sim 50$~mK.

The strongly anisotropic NFL behavior observed in CeRhSn suggests that magnetic moment involved comes from $J_z=\pm 3/2$ crystalline electric field ground state of Ce$^{3+}$.
In order to get an idea of the anisotropic response, we calculated the magnetic entropy of an isolated doublet $J_z=\pm 3/2$, well separated from the excited levels, 
and the results are shown in Fig.~S6.\cite{SM} 
Since there is no transverse component of the magnetic moment, the Zeeman splitting can occur only when $\phi_H \ne0^\circ$. Consequently,
$S(\phi_H)$ at 0.1~K exhibits a sharp peak at $\phi_H=0$, very similar to what is seen in Fig.~\ref{phase}(c).

This, of course, does not imply that the NFL behavior in CeRhSn comes from nearly free Ce$^{3+}$ ions;
entropy releasing continuously occurs below 7~K.
More significantly, the anisotropic NFL behavior collapses upon the metamagnetic crossover occurring at $\mu_0H_{\parallel a}=3$~T.
This never happens for the isolated $J_z=\pm 3/2$ doublet well separated from excited levels.
It is likely that the metamagnetic crossover is of collective origin.

Regarding the metamagnetic crossover, the possibility of a local moment metamagnetism is proposed in Ref. \ref{Tokiwa2015SA}, 
which is analogous to a spin flop along the Ising axis of an antiferromagnet.
This scenario, which assumes the Ising axis along the $a$ axis, is unlikely to be the case since no symmetry breaking of the hexagonal lattice has been reported.
Moreover, the fact that  $\chi_a(T) < \chi_c(T)$ at any temperature implies that the $a$ axis is not the magnetically easy axis.
Indeed, the magnetization in $H \parallel a$ reaches less than half of that in $H \parallel c$, even above $H_{\rm m}$ [Fig.~\ref{MH}(b)].

Even if dominant spin correlations were ferromagnetic type, a spin reorientation from the easy- to hard-magnetization axes is unlikely to occur 
because CeRhSn is a paramagnet in which local moments would not be pinned strongly enough.
A possible scenario is an occurrence of a Lifshitz transition resulting in the metamagnetic crossover and 
the suppression of key spin fluctuations for the NFL behavior.
Further investigations are needed to resolve remaining puzzles in CeRhSn.
NMR measurements under a precisely-aligned in-plane magnetic field may provide useful information.

In general, quantum criticality develops only near a QCP.
In CeRhSn, however, the NFL behavior survives over a wide range of tuning parameter $H_{\parallel a}$ values.
This feature resembles the enigmatic quantum critical ``phases'' proposed for $\beta$-YbAlB$_4$ and MnSi,\cite{Matsumoto2011Science,Tomita2015Science,Doiron-Leyraud2003Nature}
in which quantum criticality was observed in a wide range of hydrostatic pressures, even far away from the phase boundary of the magnetic order.
Therefore, the NFL nature in CeRhSn is not associated with a conventional magnetic QCP.
Moreover, in the $T$--$H_{\parallel a}$ map in Fig.~\ref{phase}(d), there are high and moderate entropy regimes, separated around $\mu_0H_{\parallel a} \sim 0.5$~T.
Around this characteristic field ($\mu_0H_{\parallel a}\sim0.5$~T), 
a magnetic phase transition is induced by in-plane stress,\cite{Kuchler2017PRB}
supporting a possible change in ground states.
In other words, the entropy plateau [Fig.~\ref{Field}(b)] indicates a partial lift of the degenerate ground states at this field.
This two-step decrease of the entropy is accompanied by a qualitative change in the exponent $n$ of $C(T)$ (Table~\ref{Table}).

\section{Summary}
We have found that a magnetic field angle is a powerful tuning parameter for the NFL nature in CeRhSn,
which survives up to a relatively high magnetic field when the field is applied precisely along the quasi-kagome plane.
The multidimensional entropy landscape and the field-angle-resolved magnetization measurements reveal that
the NFL behavior in CeRhSn is associated with Ising spin fluctuations along the $c$ axis.
This strong Ising anisotropy is likely to reflect the $J_z=\pm 3/2$ crystalline electric field ground state of Ce$^{3+}$,
although concrete mechanisms of the NFL nature and two-step decrease of the entropy under a weak transverse magnetic field remain unresolved.
These features may capture new aspects of the strongly-correlated electrons system.

\begin{acknowledgments}
We acknowledge helpful discussions with H. Tsunetsugu.
Supported by a Grant-in-Aid for Scientific Research on Innovative Areas ``J-Physics'' (JP15H05883, JP16H01076, JP18H04306) from MEXT, and 
JSPS KAKENHI Numbers JP17K05545, JP18H01161, and JP18H01164.
\end{acknowledgments}

\clearpage
\onecolumn
\renewcommand{\thefigure}{S\arabic{figure}}
\setcounter{figure}{0}

\small
\begin{center}
{\large  Supplemental Material for \\
\vspace{-0.04in}
\textbf{Field-angle-resolved landscape of non-Fermi-liquid behavior\\ in the quasi-kagome Kondo lattice CeRhSn}}\\
\vspace{0.1in}
\normalsize
Shunichiro Kittaka,$^{1,2}$ Yohei Kono,$^{1,2}$ Suguru Tsuda,$^{3}$\\ Toshiro Takabatake,$^{3}$ Toshiro Sakakibara$^{2}$\\
\textit{$^1$Department of Physics, Faculty of Science and Engineering, Chuo University, Kasuga, Bunkyo-ku, Tokyo 112-8551, Japan}\\
\textit{$^2$Institute for Solid State Physics, University of Tokyo, Kashiwa, Chiba 277-8581 \\}
\textit{$^3$Quantum Matter Program, Graduate School of Advanced Science and Engineering, Hiroshima University, Higashi-Hiroshima 739-8530, Japan}\\
\end{center}

\normalsize
\vspace{0.05in}
\section*{Magnetization measurements}
\vspace{0.05in}

In this study, 
dc magnetization was measured using a home-made capacitively detected Faraday magnetometer combined with a two-axis rotation device.\cite{Nakamura2018JPSJ2}
The angle $\phi_H$ between $H$ and the $ab$ plane was controlled precisely by rotating the sample around the $a^\ast$ axis using the tilting stage of the two-axis rotation device.
The piezo-stepper-driven goniometer attached on the tilting stage was fixed so that the $a^\ast$ axis was nearly parallel to the rotational axis of the tilting stage.

First, we carefully placed the field orientation exactly parallel to the $ab$ plane via magnetic-torque measurements.
Magnetic torque can be detected as a capacitance change of a capacitor transducer under a zero field gradient.
Figure~S1(a) shows the relative change in the raw capacitance data $\Delta C$ with $H$, measured under a zero field gradient for several $\phi_H$ values.
By rotating the tilting stage, a large dip in $\Delta C(H)$ around $\mu_0H \sim 3$~T, originating from the metamagnetic crossover, 
was suppressed and changed into a peak.
This sign change of the magnetic-torque component in $\Delta C$ indicates that $H$ crossed the $ab$ plane.
Therefore, we defined the angle at which $\Delta C(H)$ became nearly independent of $H$ as $\phi_H=0^\circ$ [dashed line in Fig.~S1(a)].
Owing to this fine-tuning, the magnetic field could be applied parallel to the $ab$ plane with a high precision, better than $1^\circ$.

Then, using the same setup, we performed magnetization measurements.
Figure~S1(b) shows the magnetization curve $M(H)$ for $\phi_H=1^\circ$ at 0.2~K, compared with $M(H)$ for $\phi_H=0^\circ$ at 0.15~K.
Above the metamagnetic-crossover field $H_{\rm m}$, no anomaly was found up to 14~T.
From this figure, it is evident that the $M(H)$ curve is not affected significantly 
by a possible tiny field misalignment less than $1^\circ$, as well as a slight difference in temperature.

\vspace{0.1in}
\section*{Calorimetric measurements}
\vspace{0.05in}

Figure S2 represents the raw data of the measurements for evaluating the field-rotational magnetocaloric effect.
We precisely measured the relative change in the sample temperature, $\Delta T(\phi_H)=T(\phi_H)-T_0$, upon a small-angle ($\Delta\phi_H$) rotation of the magnetic field
with a rotational speed of $d\phi/dt=25$~sec/deg.
Here, $T_0$ is the base temperature of the sample and $\Delta\phi_H$ is set to $1^\circ$ or $0.5^\circ$.
The magnetocaloric effect, i.e., $dT/d\phi_H$, was evaluated from the initial slope of $\Delta T(\phi_H)$ just after starting the field rotation.
We confirmed that the initial slope corresponds well between the clockwise and anti-clockwise field-rotation data (inset of Fig. S2).
This fact evidences that the extrinsic heat-transfer effect is negligibly small,\cite{Kittaka2018JPSJ2} and strengthens the reliability of the present measurements.
In a similar way, as exemplified in Fig.~S3, the conventional field-sweep magnetocaloric effect, i.e., $dT/dH$, was evaluated from the initial slope of $\Delta T(H)=T(H)-T_0$, 
which was measured with a field sweep rate of $dH/dt=0.2$~mT/sec for $H \parallel a$ (0.2 and 1 mT/sec below and above 0.3 T, respectively, for $H \parallel c$).

The field-angle-resolved landscapes of the thermodynamic quantities shown in Figs. 5(a)--5(c) were constructed using the data in Figs.~S4(a)--S4(c), respectively.
As shown in Fig.~S4(a), 
the low-field $C(\phi_H)/T$ measured at a finite temperature of 0.1~K shows peaks at non-zero $\phi_H$ because of the development of a Schottky-type anomaly by applying $H_{\parallel c}$, 
which results in a prominent dip at $\phi_H=0^\circ$.
By contrast, $S(\phi_H)$ always shows a peak centered at $\phi_H=0^\circ$, directly reflecting the high degeneracy of the ground states [Fig.~S4(c)].
Thus, entropy is more powerful to reveal the field-angle dependence of quantum criticality than specific heat.

According to the Maxwell relation, we obtain $(\partial S/\partial \phi_H)_{H,T}=(\partial \tau_\phi/\partial T)_{H,\phi_H}$, where $\tau_\phi$ denotes the in-plane magnetic torque.
By taking advantage of this relation, we can define symmetric crystalline axes along which $H$ induces $\tau_\phi=0$. 
For CeRhSn, $\phi_H=0^\circ$ can be defined with high precision much better than $0.5^\circ$ 
because the field-rotational magnetocaloric effect $dT/d\phi_H \propto (\partial \tau_\phi/\partial T)_{H,\phi_H}$ changes its sign sharply at $\phi_H=0^\circ$ [Fig.~S4(b)].
It also gives a good indication of the angle range where the entropy changes drastically.
The contour maps in Figs.~5(a)--5(c) help us confirm these features more easily. 

Figures~S5(a) and S5(b) show the raw data of $S/\gamma T$, 
which were used for constructing the $T$--$H_{\parallel a}$ and $T$--$H_{\parallel c}$ landscapes in Fig. 5(d), respectively. 
Although $S/\gamma T$ decreases rapidly and monotonically in $H \parallel c$, 
it shows a two-step decrease in $H \parallel a$.
The first, low-field decrease around 0.5~T is gradually smeared out with increasing temperature.
The second decrease around 3~T shifts toward higher fields with increasing temperature, following the metamagnetic crossover field $H_{\rm m}(T)$.
Uncommonly, there is a wide field region, where $S/\gamma T$ is insensitive to a tuning parameter $H_{\parallel a}$, below $H_{\rm m}$.

\vspace{0.1in}
\section*{Model calculation}
\vspace{0.05in}

The specific heat of an isolated doublet $J_z=\pm 3/2$ well separated from the excited levels can be calculated as
\begin{equation}
C(T,H,\theta)=k_{\rm B}N_{\rm A}\biggl\{\frac{x(T,H,\theta)}{\cosh[x(T,H,\theta)]}\biggl\}^2,
\end{equation}
where $x(T,H,\theta)=(3/2)g_J\mu_{\rm B}H\cos\theta/(k_{\rm B}T)$,
$\mu_{\rm B}$ is the Bohr magneton, $k_{\rm B}$ is the Boltzmann constant, $N_{\rm A}$ is the Avogadro constant, and $g_J=6/7$.
The magnetic field angle $\theta$ is measured from the $z$ direction.
By using this equation, the field-angle-resolved landscape of the entropy
\begin{equation}
S(T,H,\theta)=\int_0^T \frac{C(T^\prime,H,\theta)}{T^\prime}dT^\prime
\end{equation}
can be calculated and the results for $T=0.1$~K are shown in Fig.~S6. 
A sharp peak of $S(\phi_H)$ at $\phi_H=0^\circ$, seen in Fig.~5(c), can be explained qualitatively by using this simple model.

\appendix
\renewcommand{\thefigure}{S\arabic{figure}}
\renewcommand{\thetable}{S\arabic{table}}
\setcounter{figure}{0}

\clearpage
\begin{figure}
\begin{center}
\includegraphics[width=4.3in]{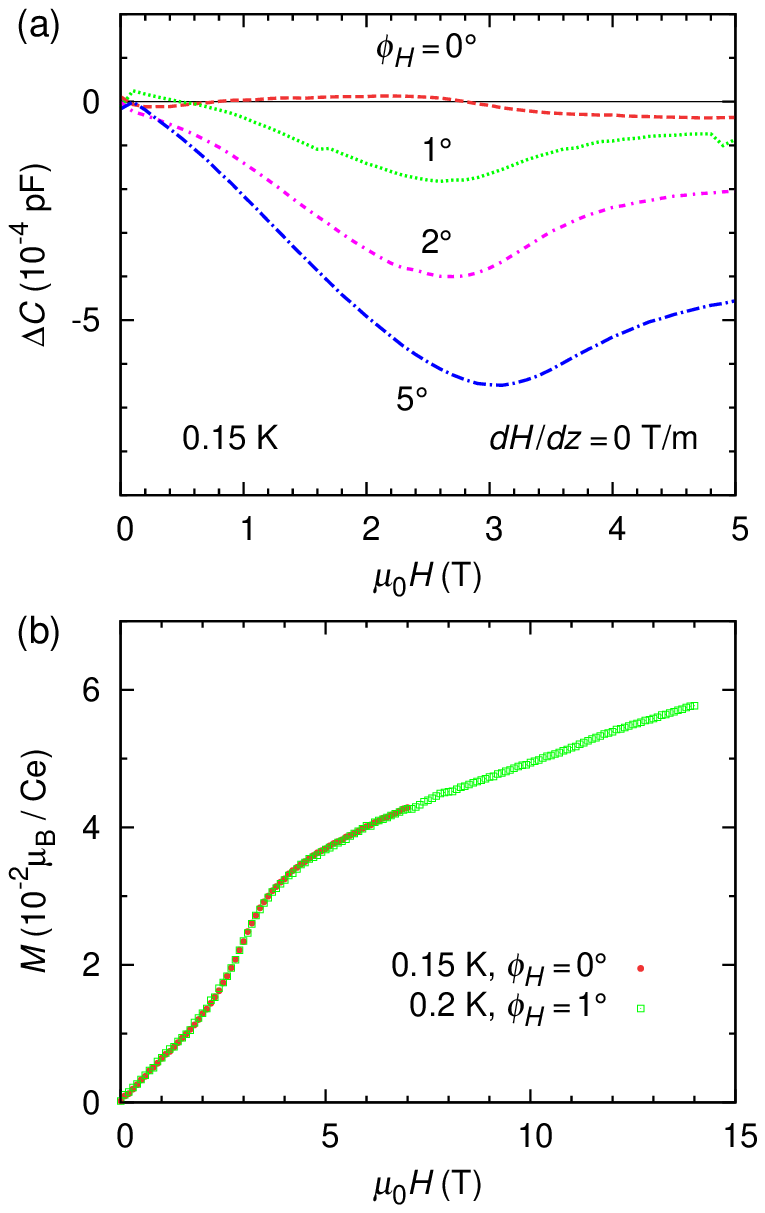} 
\end{center}
\caption{
(a) Relative change in the raw capacitance data, $\Delta C$, measured with a zero field gradient at 0.15 K for several $\phi_H$.
The solid line is the zero-torque state.
(b) The magnetization curve $M(H)$ for $\phi_H=1^\circ$ up to 14~T measured at 0.2 K, compared with $M(H)$ for $\phi_H=0^\circ$ measured at 0.15~K.
}
\end{figure}

\clearpage
\begin{figure}
\begin{center}
\includegraphics[width=5.in]{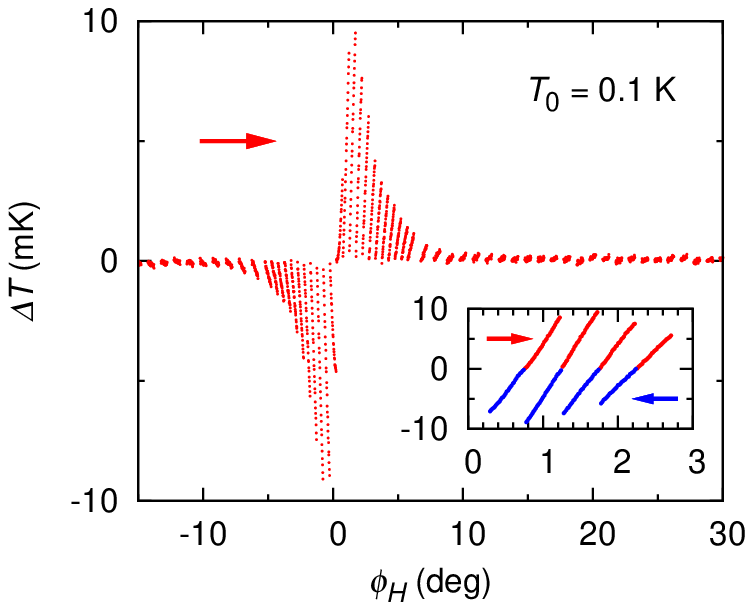} 
\end{center}
\caption{
Relative change in the sample temperature, $\Delta T(\phi_H)=T(\phi_H)-T_0$,
in response to a quasi-adiabatic rotation of an externally applied magnetic field of 2~T by $1^\circ$ or $0.5^\circ$ at $T_0=0.1$ K.
The magnetic field was rotated after the sample temperature became stable at $T_0$.
The inset compares the same data (red; the clockwise field-rotation data) with the anti-clockwise field-rotation data (blue), in the vicinity of $\phi_H=0^\circ$.
}
\end{figure}

\clearpage
\begin{figure}
\begin{center}
\includegraphics[width=5.in]{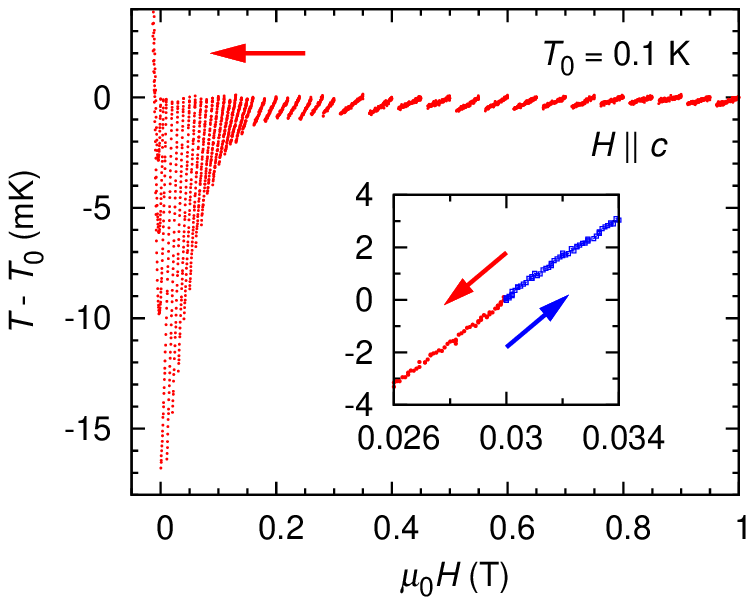} 
\end{center}
\caption{
Relative change in the sample temperature, $\Delta T(H)=T(H)-T_0$,
in response to a quasi-adiabatic change in the external magnetic-field strength at $T_0=0.1$ K for $H \parallel c$.
The magnetic field was changed after the sample temperature became stable at $T_0$.
The inset compares between the field increasing and decreasing data around $\mu_0H=0.03$~T.
}
\end{figure}

\clearpage
\begin{figure}
\begin{center}
\includegraphics[width=5.in]{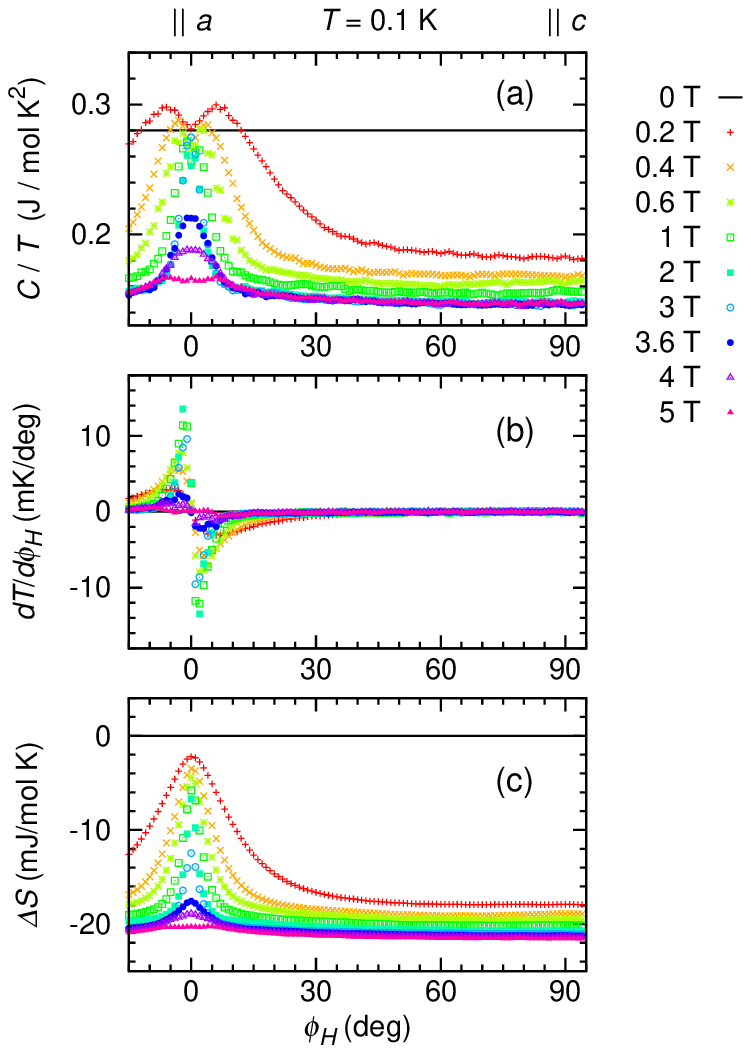} 
\end{center}
\caption{
Field-angle $\phi_H$ dependence of (a) the specific-heat data $C(H,\phi_H)/T$,
(b) rotational magnetocaloric effect $dT/d\phi_H(H,\phi_H)$, and
(c) entropy change $\Delta S(H,\phi_H) =\Delta S_\phi(H,\phi_H)+\Delta S_H(H,90^\circ)$ [$=S(H,\phi_H)-S(H=0)$] at 0.1~K.
There data were used for constructing the entropy landscape shown in Figs.~5(a)--5(c).
}
\end{figure}

\clearpage
\begin{figure}
\begin{center}
\includegraphics[width=5.in]{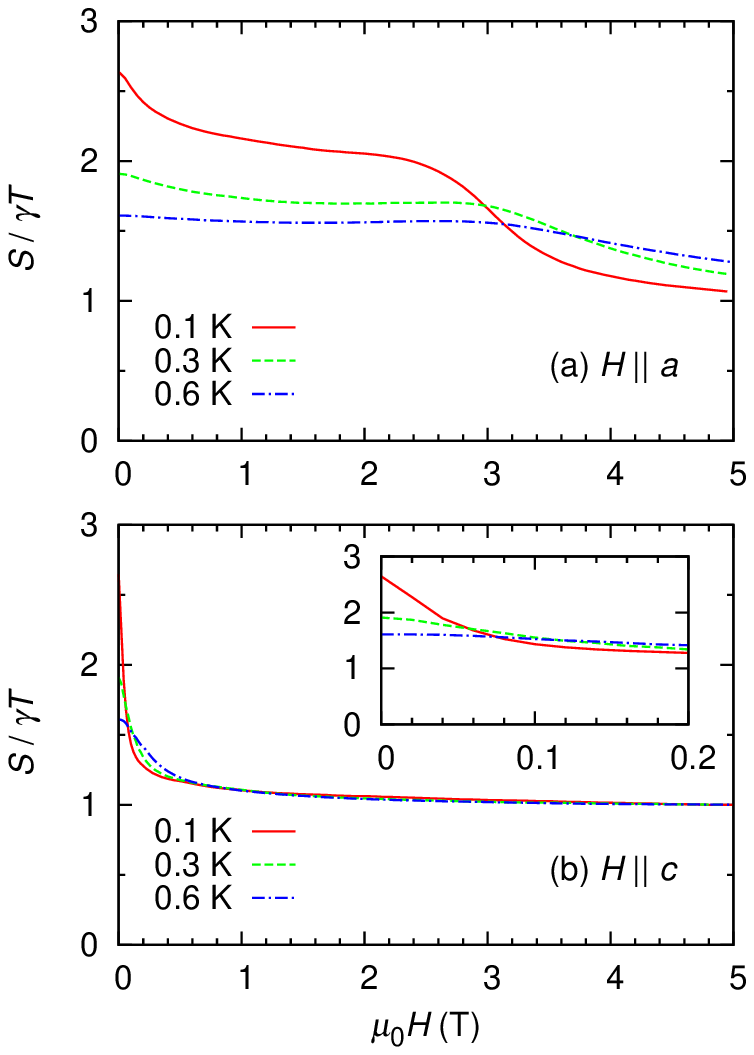} 
\end{center}
\caption{
Field dependence of (a) $S/\gamma T$ at $\phi_H=0^\circ$ ($H \parallel a$) and (b) $\phi_H=90^\circ$ ($H \parallel c$).
Inset in (b) is an enlarged view in the low-field region for $H \parallel c$.
These data were used for constructing the entropy landscape shown in Fig.~5(d).
}
\end{figure}

\clearpage
\begin{figure}
\begin{center}
\includegraphics[width=5.in]{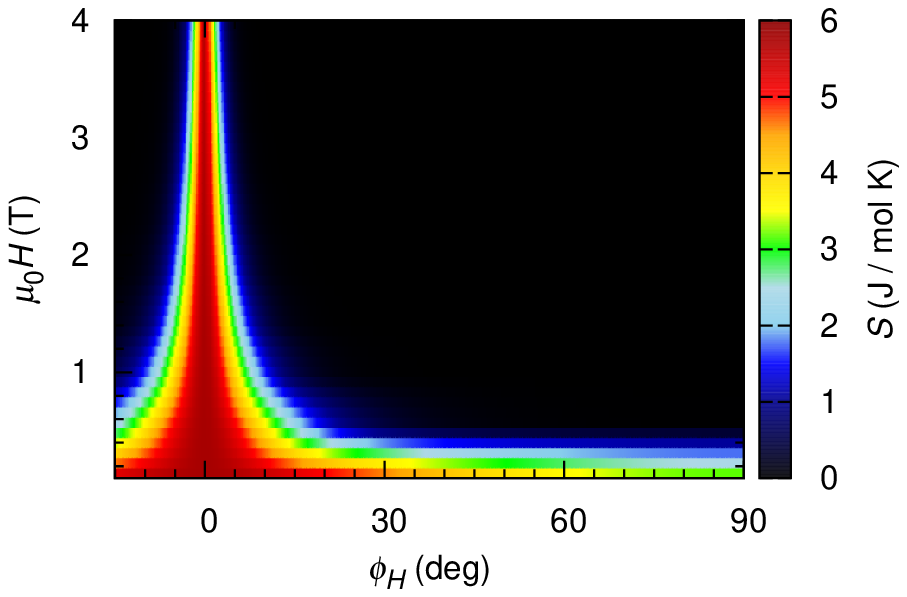} 
\end{center}
\caption{
Field-angle-resolved landscape of the entropy $S(H,\phi_H)$ at $T=0.1$~K calculated by using a simple model for an isolated doublet $J_z=\pm 3/2$.
Here, $\phi_H=\pi/2-\theta$.
}
\end{figure}

\end{document}